\documentclass[aps,showpacs,twocolumn]{revtex4-1}
\usepackage{latexsym,epsfig,graphicx}

\begin{document}

\title{Stability Analysis and Area Spectrum of 3-Dimensional Lifshitz Black Holes}
\author{Bertha Cuadros-Melgar}
\affiliation{Department of Physics, National Technical University of
Athens, Zografou Campus GR 157 73, Athens, Greece}
\affiliation{Physics Department, University of Buenos Aires
FCEN-UBA and IFIBA-CONICET, Ciudad Universitaria, Pabell\'on 1, 1428, Buenos Aires, Argentina}

\author{Jeferson de Oliveira}
\affiliation{Instituto de F\'\i sica, Universidade de S\~ao Paulo, C.P.66318, CEP 05315-970, S\~ao Paulo, Brazil}

\author{C. E. Pellicer}
\affiliation{Instituto de F\'\i sica, Universidade de S\~ao Paulo, C.P.66318, CEP 05315-970, S\~ao Paulo, Brazil}

\begin{abstract}
In this work, we probe the stability of a $z=3$ three-dimensional Lifshitz
black hole by using scalar and spinorial perturbations. 
We found an analytical expression for the quasinormal
frequencies of the scalar probe field, which perfectly agree with the
behavior of the quasinormal modes obtained numerically. 
The results for the numerical analysis of the spinorial perturbations 
reinforce the conclusion of the scalar analysis, {\it i.e.}, the model is stable under scalar and spinor perturbations. As an application we found the 
area spectrum of the Lifshitz black hole, which turns out to be equally spaced.
\end{abstract}


\maketitle

\section{Introduction}

Some decades ago Regge and Wheeler began a pioneering study of a small perturbation in the background of a black hole in order to get information of the stability of this object~\cite{rg}, a problem that was continued by Zerilli~\cite{zerilli}. The oscillations found in these studies are not normal modes due to the emission of gravitational waves; thus, their frequencies are complex and, as a result, the oscillations are damped. 

The terminology quasinormal mode (QNM) and quasinormal frequency (QNF), aiming to name these new modes and their frequencies, was first pointed out by Vishveshwara~\cite{vivesh} and Press~\cite{press}. Although initially studied in black hole backgrounds, the concept of QNM applies to a much broader class of systems.
The QNMs of black holes were first numerically calculated by Chandrasekhar and Detweiler~\cite{chandra} showing that the amplitude is dominated by a ringing characteristic signal at intermediate times. The QNMs are independent of the particular initial perturbation that excited them. The frequencies and damping times of the oscillations depend only on the parameters of the black hole and are, therefore, the ``footprints'' of this structure. 
Soon, the connection of QNMs to astrophysics was established by noting
that their existence can lead to the detection of black holes through
the observation of the gravitational wave spectrum. The interest in
QNMs has motivated the development of numerical and analytical
techniques for their computation (see~\cite{kok,nol,bcs} for a
review). Also,  the study of the quasinormal spectrum gives information about the stability aspects of black
hole solutions using probe classical matter fields (scalar, electromagnetic,
spinorial) evolving in the geometry
without backreacting on the spacetime background. Much has been
done in that direction, not only in four dimensions
\cite{elcio1}\cite{elcio2}, but also in two \cite{cardosoBTZ}, and in more than
four \cite{branas}.

Aside from the study of the stability of the solutions, the QNFs are
important in the context of the gauge-gravity correspondence, whose
most celebrated example is the duality between the type IIA-B string
theory in $AdS_{5}\times S^{5}$ spacetime and the four-dimensional
supersymmetric Yang-Mills theory \cite{maldacena}. Such a correspondence
can be generalized for those cases in which there is an event horizon in the 
gravity side. In this case the Hawking temperature of the black hole
is related to the temperature of a thermal field theory defined at the anti-de 
Sitter (AdS) boundary. Also, as a consequence of the correspondence, the
quasinormal spectrum corresponds to the poles of thermal Green
functions \cite{birm}, more precisely, the inverse of the imaginary part of the fundamental quasinormal frequency can be interpreted as the dual field theory relaxation time \cite{hubeny}.


Another interesting application of QNMs appears in the context of black hole thermodynamics. Some decades ago Bekenstein~\cite{bek} suggested that the horizon area of a black hole must be quantized, so that the area spectrum has the form $A_n = \gamma n \hbar$, with $\gamma$ a dimensionless constant to be determined. The first proposal to calculate this constant through QNMs was made by Hod~\cite{hod}. Accordingly, the real part of the asymptotic quasinormal mode can be seen as a transition frequency in the semiclassical limit, and its quantum emission causes a change in the mass of the black hole, which is related to the area. In this way, the constant $\gamma$ for a Schwarzschild black hole was determined as $\gamma=4 \ln 3$. Later, Kunstatter~\cite{kuns} repeated the calculation quantizing the adiabatic invariant $I=\int dE/\omega(E)$ via the Bohr-Sommerfeld quantization and using the real part of the QNF as the vibrational frequency. The result was exactly the same as Hod's. However, recently Maggiore~\cite{maggiore} pointed out that QNMs should be described as damped harmonic oscillators, thus, the imaginary part of the QNF should not be neglected, and the proper physical frequency is the module of the entire QNF. Moreover, when considering the quantization of the adiabatic invariant, the frequency to be considered is that corresponding to a transition between two neighboring quantum levels. With this identification, the constant $\gamma$ for a Schwarzschild black hole becomes $\gamma=8\pi$, a result that coincides with the value calculated by other methods~\cite{othmet}. The consequences of Hod's and Maggiore's proposals were promptly studied in several spacetimes~\cite{pap,magg}.


In this paper, we are interested in the study of the stability of the $z=3$, three-dimensional Lifshitz black hole found in the context of the so-called new massive gravity (NMG)~\cite{aggh}. Moreover, as an application of our QNM results we aim to calculate the area spectrum of this black hole.

NMG is a novel parity-preserving, unitary~\cite{unit}, power-counting super-renormalizable~\cite{ren}, three-dimensional model describing the propagation of a massive positive-energy graviton with two polarization states of helicity $\pm 2$ in a Minkowski vacuum, whose linearized version is equivalent to the Pauli-Fierz theory for a massive spin-2 field in three dimensions.
The action of NMG consists of a ``wrong sign'' Einstein-Hilbert term plus a quadratic curvature term given by a precise combination of the Ricci tensor and the curvature scalar, which introduces a mass parameter~\cite{nmg}.
As with other massive gravity theories, NMG also admits black hole-type solutions with several asymptotics and additional parameters~\cite{nmg2,bhnmg}. Even though this last feature could challenge the usual Einstein-Hilbert gravity, it is seen that the definition of mass in this new type of black holes is a conserved charge computed from a combination of the black hole parameters, which satisfies the first law of thermodynamics. A study of QNMs in these static new type of black holes has been performed in~\cite{kwon}.

The black holes we take into account for our study are asymptotically
Lifshitz, {\it i.e.}, they exhibit the anisotropic scale invariance,
$t\rightarrow \lambda^z t$, $\vec x \rightarrow \lambda\vec x$, where 
$z$ is the dynamical critical exponent. Specifically, we deal with the
solutions found for the particular case of $z=3$ and a precise value of
the mass parameter~\cite{aggh}. The general class of these solutions
are important in the context of gauge-gravity
duality~\cite{kachru,alt} and were also investigated in other
background theories~\cite{lifBH1,lifBH2,lifBH3,lifBH4}. No stability study 
of black holes with Lifshitz asymptotics in three dimensions in
the scenario of NMG has been performed yet. We aim to give some
contribution to this issue by considering the QNF of scalar and
spinorial matter fields in the probe limit, {\it{i.e}}, there are no
backreaction effects upon the asymptotically Lifshitz black hole metric. Spinor
fields have been extensively studied in general relativity
\cite{finster1}\cite{finster2}, and their quasinormal frequencies have also
been considered \cite{cho2}\cite{cho1}\cite{owenjef}.   

The paper is organized as follows. In section 2 we introduce the
Lifshitz black holes and discuss their causal structure. Sections 3
and 4 are dedicated to the study of stability under scalar and
spinorial perturbations with special emphasis on the massless spinor for the latter. In section 5 we present the numerical analysis for both kinds of perturbations showing the QNMs and the corresponding QNFs computed in each case. Section 6 is devoted to the calculation of the area spectrum of these black holes as an application of our quasinormal spectrum. Finally, we discuss our results and conclude in section 7.

\section{Lifshitz Black Holes in Three Dimensions}

In this section we review the black hole solutions we will consider within this paper, and we comment some of their features.

The NMG theory \cite{nmg} is defined by the $(2+1)$-dimensional action,
\begin{equation}\label{action}
S=\frac{1}{16\pi G} \int d^3 x \sqrt{-g} \,\left[R-2\lambda-\frac{1}{m^2}\left(R_{\mu\nu}R^{\mu\nu}-\frac{3}{8}R^2\right)\right]\,,
\end{equation}
where $m$ is the so-called ``relative'' mass parameter, and $\lambda$ is the
three-dimensional cosmological constant. Defining the dimensionless
parameters, $y=m^2 l^2$ and $w=\lambda l^2$, it is found that the theory
exhibits special properties at the points $y=\pm 1/2$. When looking
for black hole solutions with Lifshitz asymptotics, it is precisely at
the point $y=-1/2$, $w=-13/2$, with Lifshitz scaling $z=3$, where the
field equations turn out to be solved by \cite{aggh}
\begin{equation}\label{metric}
ds^{2}=-a(r)\frac{\Delta}{r^{2}}dt^{2}+\frac{r^{2}}{\Delta}dr^{2}+r^{2}d\phi^{2}\,,
\end{equation}
where 
\begin{equation}\label{a}
a(r)=\frac{r^{4}}{l^{4}}\,,
\end{equation}
and
\begin{equation}\label{delta}
\Delta=-Mr^{2}+\frac{r^{4}}{l^{2}}\,,
\end{equation}
with $M$ an integration constant and
$l^{2}=-\frac{13}{2\lambda}$. Also, the NMG admits as a solution, the
well-known Ba\~nados-Teitelboim-Zanelli (BTZ) black hole with the dynamical critical exponent $z=1$. 
As we shall see below in more detail, this metric (\ref{metric}) exhibits a regular 
single event horizon located at $r=r_{+}=l\sqrt{M}$ and a spacetime 
singularity at $r=0$. Besides, the surface $r=r_{+}$ acts as a one-way membrane
for physical objects as we can see from the norm of a  vector $\chi$
normal to a given surface $s$. Since $s$ has to be null in order to be a
one-way membrane, the norm of $\chi$ must be null as well,
{\it i.e.}, $g^{rr}=0$, which occurs at $r=r_{+}$.

From the behavior of the Kretschmann invariant for the metric
(\ref{metric}),

\begin{equation}\label{kresh}
R_{\mu\nu\lambda\sigma}R^{\mu\nu\lambda\sigma}= -\frac{4}{l^{4}r^{4}}\left[8r_{+}^{4}-48r_{+}^{2}r^{2}+91r^{4}\right]\,,
\end{equation}
we see that, for $r\rightarrow r_{+}$
\begin{equation}\label{kresh_h}
R_{\mu\nu\lambda\sigma}R^{\mu\nu\lambda\sigma}\rightarrow -\frac{204}{l^{4}}\,,
\end{equation}
 and for $r \rightarrow 0$
\begin{equation}\label{kresh_sing}
R_{\mu\nu\lambda\sigma}R^{\mu\nu\lambda\sigma}\rightarrow \infty\,.
\end{equation}
Thus, the black hole solution has a genuine spacetime singularity at
the origin $r=0$ and an event horizon at $r=r_{+}$. Nevertheless, to
see if the singularity is timelike, spacelike, or null we have to
construct the Penrose-Carter diagram. First of all, we must remove
the coordinate singularity at $r=r_{+}$. Rewriting the metric in terms
of null coordinates $(U,V)$
\begin{equation}\label{null2}
U=e^{r_{+}^{3}(t+r_{*})},\hspace{0.3cm}V=-e^{-r_{+}^{3}(t-r_{*})}\,.
\end{equation}

where $r_{*}$ is the tortoise coordinate shown in the next section, we get

\begin{equation}\label{metric_null2}
ds^{2}=-\frac{1}{4}\left(\frac{r}{r_{+}}\right)^{6}\left(1+\frac{r_+}{r} \right)^2e^{-\frac{2r_{+}}{r}}dUdV\,,
\end{equation}
which is manifestly regular at $r=r_{+}$. 

\begin{figure}[htb!]
\begin{center}
\includegraphics[height=8cm]{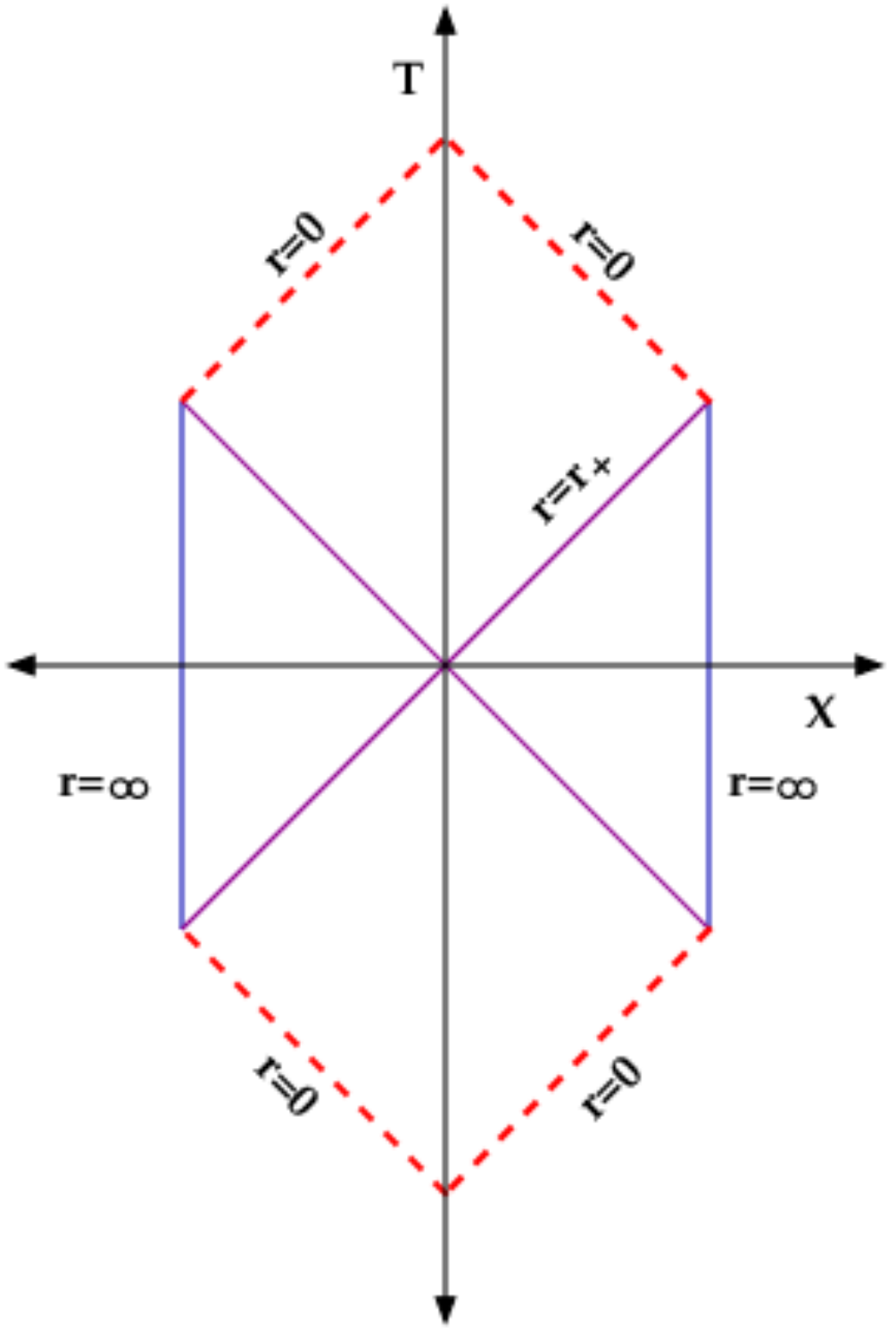}
\end{center}
\caption{Penrose-Carter diagram for the Lifshitz black hole. The singularity at $r=0$ is light-like and covered by a regular event horizon $r_+$.}
\label{penrose}
\end{figure}

Finally, to construct the Penrose-Carter diagram (Fig.\ref{penrose}) we use the following set of null coordinates
\begin{equation}\label{null3}
T=\tilde{U}+\tilde{V},\hspace{0.3cm}X=\tilde{U}-\tilde{V}\,,
\end{equation}
with $\tilde{U}=\arctan(U)$ and $\tilde{V}=\arctan(V)$. 
 
From this diagram we see that the spacetime singularity
is located at $r=0$, as previously observed from the behavior of the
Kretschmann invariant. Moreover, it is light-like and covered by a regular
event horizon at $r=r_{+}$.

\bigskip

\section{Scalar Perturbation}

In this section, we analyze the behavior of a scalar field perturbation
in the background of a three-dimensional Lifshitz black hole.

The scalar field obeys the Klein-Gordon equation,
\begin{equation}\label{kg}
\Box \Phi = \frac{1}{\sqrt{-g}} \partial_M\left(\sqrt{-g}g^{MN} \partial_N \right)\Phi = m^2 \Phi \,,
\end{equation}
where $m$ is the mass of the field $\Phi$. Performing the decomposition

\begin{equation}
\Phi (t,r,\phi) = \Psi(t,r)\,e^{i\kappa\phi},
\end{equation}

The Klein-Gordon equation takes the form,
\begin{eqnarray}\label{kg3}
-\partial_t ^2 \Psi + \frac{r^4}{l^6} \left(1-\frac{Ml^2}{r^2}\right) \left(\frac{5r^3}{l^2}-3Mr \right) \partial_r \Psi + &&\nonumber \\
+ \frac{r^8}{l^8} \left( 1-\frac{Ml^2}{r^2}\right)^2 \partial_r ^2 \Psi - &&\nonumber \\
-\frac{r^4}{l^6}\left( m^2 r^2 + \kappa^2 \right)\left( 1-\frac{Ml^2}{r^2}\right) \Psi &=&0\,.\phantom{xx}
\end{eqnarray} 

Even though this equation has an analytical solution, as we will see in what follows, it is also useful to check the numerical results. With this goal we further decompose $\Psi=X(t,r_*)/\sqrt{r}$, where the {\it tortoise} coordinate $r_{*}$ is given by

\begin{equation}\label{tortoise}
r_* = l^4 \left[ -\frac{1}{M^{3/2}l^3}\; \hbox{arccoth} \left(\frac{r}{l\sqrt{M}}\right) + \frac{1}{Ml^2 r}\right] \,.
\end{equation}
In this way the Klein-Gordon equation reduces to

\begin{equation}\label{schrod-sc}
-\partial_t ^2 X + \partial_{r_*} ^2 X = V(r)X \,,
\end{equation}

where $V(r)$ is the scalar effective potential given by

\begin{eqnarray}\label{sc-pot}
V(r) &=& \left(\frac{7}{4l^8} + \frac{m^2}{l^6} \right) r^6 - \left(\frac{5M}{2l^6} + \frac{Mm^2}{l^4} - \frac{\kappa^2}{l^6}\right) r^4 + \nonumber \\
&&+\left( \frac{3M^2}{4l^4} - \frac{M\kappa^2}{l^4}\right) r^2\,.
\end{eqnarray}

Now let us come back to the issue of finding an exact solution for Eq.(\ref{kg3}). We set the time dependence of the field $\Psi(t, r_{*})$ as $R(r)e^{-i\omega t}$
and redefine the radial coordinate as $r=l\sqrt{M}/y$. Thus, Eq.(\ref{kg3}) turns out to be 
\begin{widetext}
\begin{equation}
\partial_y ^2 R + \frac{y^2-3}{y(1-y^2)} \partial_y R - 
-\frac{l^2}{(1-y^2)} \left[-\frac{\omega^2 y^4}{M^3 (1-y^2)} + \frac{m^2}{y^2} + \frac{\kappa^2}{Ml^2} \right] R=0 \,,
\end{equation}
whose solution is given in terms of Heun confluent functions,
\begin{eqnarray}\label{solR}
R(y) &=& C_1 y^{2+\alpha} (1-y^2)^{\beta/2}\, \hbox{HeunC}\left(0,\alpha,\beta,-\frac{\beta^2}{4},\frac{\alpha^2}{4}+\frac{\kappa^2}{4M},y^2\right) + \nonumber \\
&&+C_2 y^{2-\alpha} (1-y^2)^{\beta/2}\, \hbox{HeunC}\left(0,-\alpha,\beta,-\frac{\beta^2}{4},\frac{\alpha^2}{4}+\frac{\kappa^2}{4M},y^2\right)\,,\;
\end{eqnarray}
\end{widetext}
where $C_1$ and $C_2$ are integration constants, while $\alpha=\sqrt{4+m^2 l^2}$, and $\beta=-i\, l\omega/M^{3/2}$.

Imposing the Dirichlet condition at infinity we set $C_1=0$. In order to apply the boundary condition of in-going waves at the horizon we use the following connection formula~\cite{kazakov},
\begin{widetext}
\begin{eqnarray}
\hbox{HeunC}(0,b,c,d,e,z) &=& \frac{c_1 \Gamma(1-b)\Gamma(c)}{\Gamma(1+c+k)\Gamma(-b-k)} \hbox{HeunC}(0,c,b,-d,e+d,1-z) + \nonumber \\
&&+ \frac{c_2 \Gamma(1-b)\Gamma(-c)}{\Gamma(1-c+s)\Gamma(-b-s)} (1-z)^{-c} \hbox{HeunC}(0,-c,b,-d,e+d,1-z)\,.
\end{eqnarray}
\end{widetext}
This formula connects a solution around the singular regular point $z=0$ to the corresponding solution about the singular regular point $z=1$ of the confluent Heun equation given by
\begin{equation}
z(z-1) H'' + [(b+1)(z-1)+(c+1)z] H' + (dz-\epsilon) H=0\,.
\end{equation}
The parameters $k$ and $s$ are obtained from
\begin{eqnarray}
k^2 + (b+c+1)k -\epsilon +d/2 &=&0\,, \\
s^2 + (b-c+1)s -\epsilon +d/2 &=&0\,,
\end{eqnarray}
and $\epsilon$ is related to $e$ as
\begin{equation}
\epsilon = -\frac{bc}{2}-\frac{c}{2}-\frac{b}{2} -e\,.
\end{equation}

Thus, near $y=1$ Eq.(\ref{solR}) can be written as

\begin{eqnarray}
R(y\rightarrow 1) &\approx& \xi_1 (1-y^2)^{\beta/2} \frac{\Gamma(1+\alpha)\Gamma(\beta)}{\Gamma(\alpha-k)\Gamma(1+\beta+k)} + \phantom{xxxxx}\nonumber \\
&&+\xi_2 (1-y^2)^{-\beta/2}\frac{\Gamma(1+\alpha)\Gamma(-\beta)}{\Gamma(\alpha-s)\Gamma(1-\beta+s)} \,,
\end{eqnarray}
with $\xi_i$ as constants.
As we are looking for quasinormal frequencies with negative imaginary parts, which give stable solutions, we find that for $\beta<0$ we need $\Gamma(1+\beta+k)\rightarrow \infty$. Thus, the quasinormal frequencies are
\begin{widetext}
\begin{equation}\label{freq_analitico}
\omega = 2i \frac{M^{3/2}}{l} \left[ 1+2N+\sqrt{4+m^2 l^2} - \sqrt{7+\frac{3}{2}m^2 l^2 +\frac{\kappa^2}{2M} + (3+6N)\sqrt{4+m^2 l^2}+6N(N+1)}\right]\,,
\end{equation}
\end{widetext}
where $N$ is a positive integer. 
The imaginary part of the fundamental frequency ($N=0$) is negative provided that
\begin{equation}\label{condition}
\sqrt{7+\frac{3}{2} m^2 l^2 + \frac{\kappa^2}{2M}+3\sqrt{4+m^2 l^2}} > 1+\sqrt{4+m^2 l^2} \,.
\end{equation}
While the asymptotic frequency ($N\rightarrow \infty$) is given by
\begin{equation}\label{asymptotic}
\omega_\infty = -2(\sqrt{6}-2)\,i\, \frac{M^{3/2}}{l} N <0\,.
\end{equation}
Thus, since the imaginary part of the quasinormal frequencies is negative provided that the parameters respect the relation (\ref{condition}), we can conclude that the model is stable under scalar perturbations.


\section{Spinorial Perturbation}

In this section, we are going to consider a spinorial field as a
perturbation in the spacetime given by the three-dimensional Lifshitz
black hole. We analyze the covariant Dirac equation for a two component spinor
field $\Psi$ with mass $\mu_{s}$. This equation is given by

\begin{equation}\label{dirac}
i\gamma^{(a)}e_{(a)}^{\phantom{(a)}\mu}\nabla_{\mu}\Psi-\mu_{s}\Psi=0\,,
\end{equation}
where Greek indices refer to spacetime coordinates $(t,r,\phi)$, and
the Latin indices enclosed in parentheses describe the flat tangent 
space in which the triad basis $e_{(a)}^{\phantom{(a)}\mu}$ is defined. 
The spinor covariant derivative $\nabla_{\mu}$ is given by 

\begin{equation}\label{covariant_dev}
\nabla_{\mu}=\partial_{\mu}+\frac{1}{8}\omega_{\mu}^{\phantom{\mu}(a)(b)}\left[\gamma_{(a)},\gamma_{(b)}\right]\,,
\end{equation}
where $\omega_{\mu}^{\phantom{\mu}(a)(b)}$ is the spin connection,
which can be written in terms of the triad
$e_{(a)}^{\phantom{(a)}\mu}$ as 

\begin{equation}\label{connection}
 \omega_{\mu}^{\phantom{\mu}(a)(b)}=e_{\nu}^{\phantom{\mu}(a)}\partial_{\mu}e^{(b)\nu}+e_{\nu}^{\phantom{\nu}(a)}\Gamma^{\nu}_{\phantom{\nu}\mu\sigma}e^{\sigma
 (b)}\,,
\end{equation}
where $\Gamma^{\nu}_{\phantom{\nu}\mu\sigma}$ are the metric
connections. The $\gamma^{(a)}$ denotes the usual flat gamma matrices,
which can be taken in terms of the Pauli ones. In this work we take
$\gamma^{(0)}=i\sigma_{2}$, $\gamma^{(1)}=\sigma_{1}$, and $\gamma^{(2)}=\sigma_{3}$. 

We can write the triad basis $e_{(a)}^{\phantom{(a)}\mu}$ for the metric
(\ref{metric}) as follows:

\begin{eqnarray}\label{tryad}
&e_{0}^{\phantom{0}(a)}=\frac{\sqrt{a(r)\Delta}}{r}\delta^{\phantom{0}(a)}_{0},\quad
e_{1}^{\phantom{0}(a)}=\frac{r}{\sqrt{\Delta}}\delta^{\phantom{1}(a)}_{1},&\nonumber \\ \nonumber \\
&e_{2}^{\phantom{0}(a)}=r\delta^{\phantom{2}(a)}_{2}\,,& 
\end{eqnarray}
and the metric connections,
\begin{widetext}
\begin{equation}\label{connections}\nonumber
\Gamma^{0}_{\phantom{0}01}=\frac{d}{dr}\left[\ln\left(\frac{a(r)\Delta}{r^{2}}\right)^{1/2}\right],\hspace{0.3cm}\Gamma^{1}_{\phantom{1}11}=\frac{d}{dr}\left[\ln\left(\frac{r^{2}}{\Delta}\right)^{1/2}\right],\hspace{0.3cm}
\Gamma^{1}_{\phantom{1}00}=\frac{\Delta}{2r^2}\frac{d}{dr}\left[\frac{a(r)\Delta}{r^{2}}\right],\hspace{0.3cm}\Gamma^{1}_{\phantom{1}22}=-\frac{\Delta}{r},\hspace{0.3cm}\Gamma^{2}_{\phantom{2}12}=\frac{1}{r}\,.
\end{equation}
\end{widetext}

Using these quantities it is straightforward to write down the expressions
for spin connection components. In the present case, we have
only two nonvanishing components,

\begin{eqnarray}\label{spinconnections}
\omega_{0}^{\phantom{0}(0)(1)}=\frac{1}{2\sqrt{a(r)}}\frac{d}{dr}\left(\frac{a(r)\Delta}{r^{2}}\right),\hspace{0.2cm}
\omega_{2}^{\phantom{2}(1)(2)}=-\frac{\sqrt{\Delta}}{r}\,.
\end{eqnarray}

At this point we are able to write the Dirac equation for
the two component spinor 
\begin{eqnarray}\label{two-spinor}
\Psi=\left( \begin{array}{c}
\Psi_{1}(t,r,\phi)  \\
\Psi_{2}(t,r,\phi)  \\ \end{array} \right)\,,
\end{eqnarray}
which turns to be the set of coupled  differential equations
\begin{widetext}
\begin{eqnarray}
\frac{ir}{\sqrt{a(r)\Delta}}\partial_{t}\Psi_{2}+\frac{i\sqrt{\Delta}}{r}\partial_{r}\Psi_{2}+\frac{i}{r}\partial_{\phi}\Psi_{1}+\frac{i}{4}\left[\frac{a(r)'\Delta}{a(r)r}+\frac{\Delta'}{r\sqrt{\Delta}}\right]\Psi_{2}-\mu_{s}\Psi_{1}=0\,,\label{dirac1} \\ 
-\frac{ir}{\sqrt{a(r)\Delta}}\partial_{t}\Psi_{1}+\frac{i\sqrt{\Delta}}{r}\partial_{r}\Psi_{1}-\frac{i}{r}\partial_{\phi}\Psi_{2}+\frac{i}{4}\left[\frac{a(r)'\Delta}{a(r)r}+\frac{\Delta'}{r\sqrt{\Delta}}\right]\Psi_{1}-\mu_{s}\Psi_{2}=0\,.\label{diracb}
\end{eqnarray}
\end{widetext}

In order to simplify our problem, we redefine $\Psi_1$ and $\Psi_2$ as

\begin{eqnarray}\label{redef-fields}
\Psi_{1}&=&i\left[a(r)\Delta\right]^{1/4}e^{-i\omega t+im\phi}\Phi_{+}(r),\nonumber \\
\Psi_{2}&=&\left[a(r)\Delta\right]^{1/4}e^{-i\omega t+im\phi}\Phi_{-}(r)\,,
\end{eqnarray}
and the tortoise coordinate as in the scalar case (\ref{tortoise}),

\begin{eqnarray}\label{tortoise2}
\frac{d}{dr_{*}}=\frac{\Delta\sqrt{a(r)}}{r^{2}}\frac{d}{dr}\,.
\end{eqnarray}
Thus, we can rewrite Eqs.(\ref{dirac1}) and (\ref{diracb}) as
\begin{eqnarray}
\partial_{r_{*}}\Phi_{-}-i\omega\Phi_{-}&=&i\frac{\sqrt{a(r)\Delta}}{r^{2}}\left(\hat{m}-i\mu_{s}r\right)\Phi_{+}\,,\label{dirac2}\\
\partial_{r_{*}}\Phi_{+}+i\omega\Phi_{+}&=&i\frac{\sqrt{a(r)\Delta}}{r^{2}}\left(\hat{m}+i\mu_{s}r\right)\Phi_{-}\,,\label{dirac2b}
\end{eqnarray}
where $m=i\hat{m}$.

\begin{figure}[h!]
\begin{center}
\includegraphics[width=7cm, height=9cm,angle=270]{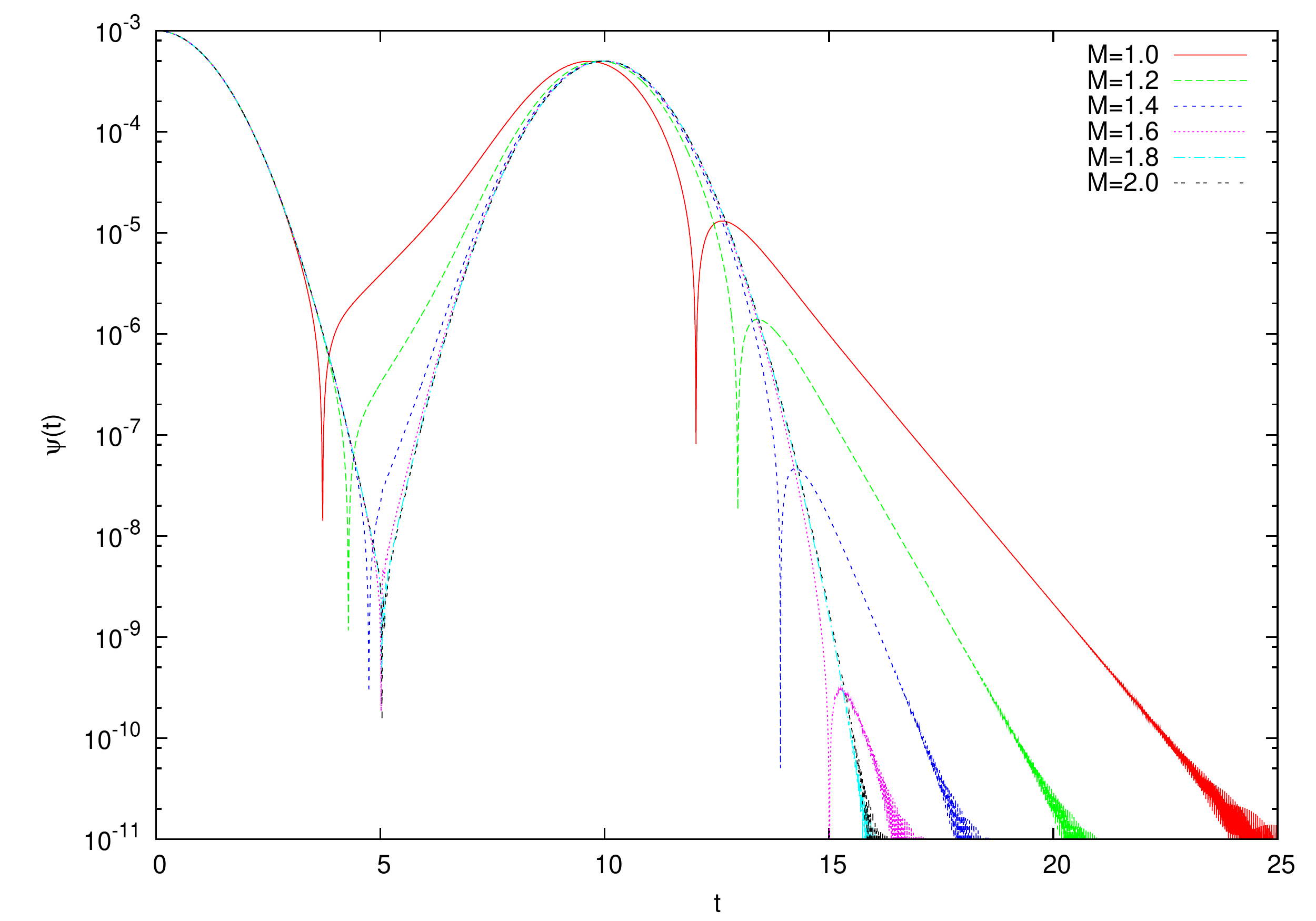}
\caption{Decay of scalar field with mass $m=1$ and $l=1$ for different
  values of black hole mass $M$.}
\label{escalar_varM}
\end{center}
\end{figure}

Furthermore, we define a new function $\theta$, a new rescaling for the spinorial components $R_{\pm}$, and a new {\it tortoise}
coordinate $\hat{r}_{*}$ through the expressions,
\begin{eqnarray}\label{relations}\nonumber
\theta&=&\arctan(\frac{\mu_{s}r}{\hat{m}})\,,\\\nonumber
\Phi_{\pm}&=&e^{\pm i\theta/2}R_{\pm}(r)\,,\\\nonumber
\hat{r}_{*}&=&r_{*}+\frac{1}{2\omega}\arctan(\frac{\mu_{s}r}{\hat{m}})\,.
\end{eqnarray}

In this way Eqs.(\ref{dirac2}) and (\ref{dirac2b}) become
\begin{equation}\label{dirac3}
\left(\partial_{\hat{r}_{*}}\pm i\omega\right)R_{\pm}=W\, R_{\mp}\,,
\end{equation}
where $W$ is the so-called superpotential,
\begin{equation}\label{superpotential}
W=\frac{i\sqrt{a(r)\Delta}\,(\hat{m}^{2}+{\mu_{s}}^{2}
  r^{2})^{3/2}}{r^{2}(\hat{m}^{2}+\mu_{s}^{2} r^{2})+\frac{\mu_{s} \hat{m}\sqrt{a(r)}\Delta}{2\omega}}\,.
\end{equation}
Notice that when $a(r)=1$, Eq.(\ref{superpotential}) reduces to the BTZ superpotential~\cite{cardoso}.

\bigskip

Finally, letting $X_{\pm}=R_{+}\pm R_{-}$ we have
\begin{equation}\label{dirac-final}
\left(\partial^{2}_{\hat{r}_{*}}+\omega^{2}\right)X_{\pm}=V_{\pm}X_{\pm}\,,
\end{equation}

where $V_\pm$ are the superpartner potentials,
\begin{equation}\label{potential1}
V_{\pm}=W^{2}\pm\frac{dW}{d\hat{r}_{*}}\,,
\end{equation}
which in the case of a massless spinor ($\mu_{s}=0$) reduces to
\begin{widetext}
\begin{eqnarray}\label{potential-weyl}
V_{\pm}=\left(-\frac{m^{2}M}{l^{4}}\mp\frac{mM}{l^{5}}\sqrt{r^{2}-Ml^{2}}\right)r^{2}
+\left(\frac{m^{2}}{l^{6}}\pm\frac{2m}{l^{7}}\sqrt{r^{2}-Ml^{2}}\right)r^{4}\,.
\end{eqnarray}
\end{widetext}

\section{Numerical Analysis}\label{numerical}

In this section, we numerically solve Eqs.(\ref{schrod-sc}) and (\ref{dirac-final}), which correspond to the scalar and massless spinorial perturbations, respectively. 
Although in the scalar case we found an analytical solution and the corresponding QNF, our motivation to perform the numerical analysis is to verify the applicability of certain numerical methods in asymptotically Lifshitz spacetimes. In particular, it would be interesting to check if the Horowitz-Hubeny method~\cite{hubeny} works well when finding the QNF.

\begin{widetext}
Using the finite difference method, we define $\psi(r_*,t) = \psi(-j \Delta r_*,l \Delta t) = \psi_{j,l}$, $V(r(r_*)) = V(-j \Delta r_*) = V_j$, and rewrite 
Eqs.(\ref{schrod-sc}) and (\ref{dirac-final}) as

\begin{equation}\label{diffin}
- \frac{\psi_{j,l+1} - 2 \psi_{j,l} + \psi_{j,l-1}}{\Delta t^2} + \frac{\psi_{j+1,l} - 2 \psi_{j,l} + \psi_{j-1,l}}{\Delta r_*^2} - V_j \psi_{j,l} + O(\Delta t^2) + O(\Delta r_*^2) = 0\,,
\end{equation}
which can be rearranged as
\begin{equation}\label{diffin_rearranged}
\psi_{j,l+1} = - \psi_{j,l-1} + \frac{\Delta t^2}{\Delta r_*^2} \left( \psi_{j+1,l} + \psi_{j-1,l} \right) + \left( 2 - 2 \frac{\Delta t^2}{\Delta r_*^2} - \Delta t^2 V_j \right)  \psi_{j,l} \,.
\end{equation}
\end{widetext}

The initial conditions $\psi(r_*,0) = f_0(r_*)$ and $\dot{\psi}(r_*,0) = f_1(r_*)$ define the values of $\psi_{j,l}$ for $l=0$ and $l=1$ and we use Eq.~(\ref{diffin_rearranged}) 
to obtain the values of $\psi_{j,l}$ for $l>1$. At $j=0$ we impose Dirichlet boundary conditions since $V(r_*)$ tends to infinity as $r_*$ tends to zero. The numerical solution 
is stable if
\begin{equation}\label{von_neumann_condition}
\frac{\Delta t^2}{\Delta r_*^2} + \frac{\Delta t^2}{4} V_{\text{MAX}}<1\,,
\end{equation}
where $V_{\text{MAX}} = V_1$ is the largest value of $V_j$ in our domain. 

This condition is verified in all cases studied here.

\bigskip

Now we are going to analyze the potential for the scalar case. 
By rewriting Eq.(\ref{sc-pot}) in terms of a new variable $z=r^2$, we obtain
\begin{widetext}
\begin{equation}\label{pot1}
V(r) = \frac{z}{l^8} \left[ \left( \frac{7}{4} + m^2 l^2 \right) z^2 - \left( \frac{5}{2} + m^2 l^2 - \frac{\kappa^2l^2}{z_h} \right) z_h z + \left( \frac{3}{4} - \frac{\kappa^2l^2}{z_h} \right) z_h^2 \right]\,,
\end{equation}
\end{widetext}
where $z_h = r_h ^2$. The parable in brackets tends to infinity as long as $\left(\frac{7}{4} +m^2l^2\right)>0$, which is consistent with the Breitenlohner-Freedman-type bound for the present case. The roots of this polynomial potential are given by
\begin{eqnarray}\label{roots}
z_0 &=&0\,,\nonumber\\
z_+ &=& z_h \,, \\
z_- &=& z_h \left[ \frac{ \frac{3}{4} - \frac{\kappa^2l^2}{z_h} }{ \frac{7}{4} + m^2l^2 } \right]\,.
\end{eqnarray}
\begin{figure}[h!]
\begin{center}
\includegraphics[width=7cm, height=9cm,angle=270]{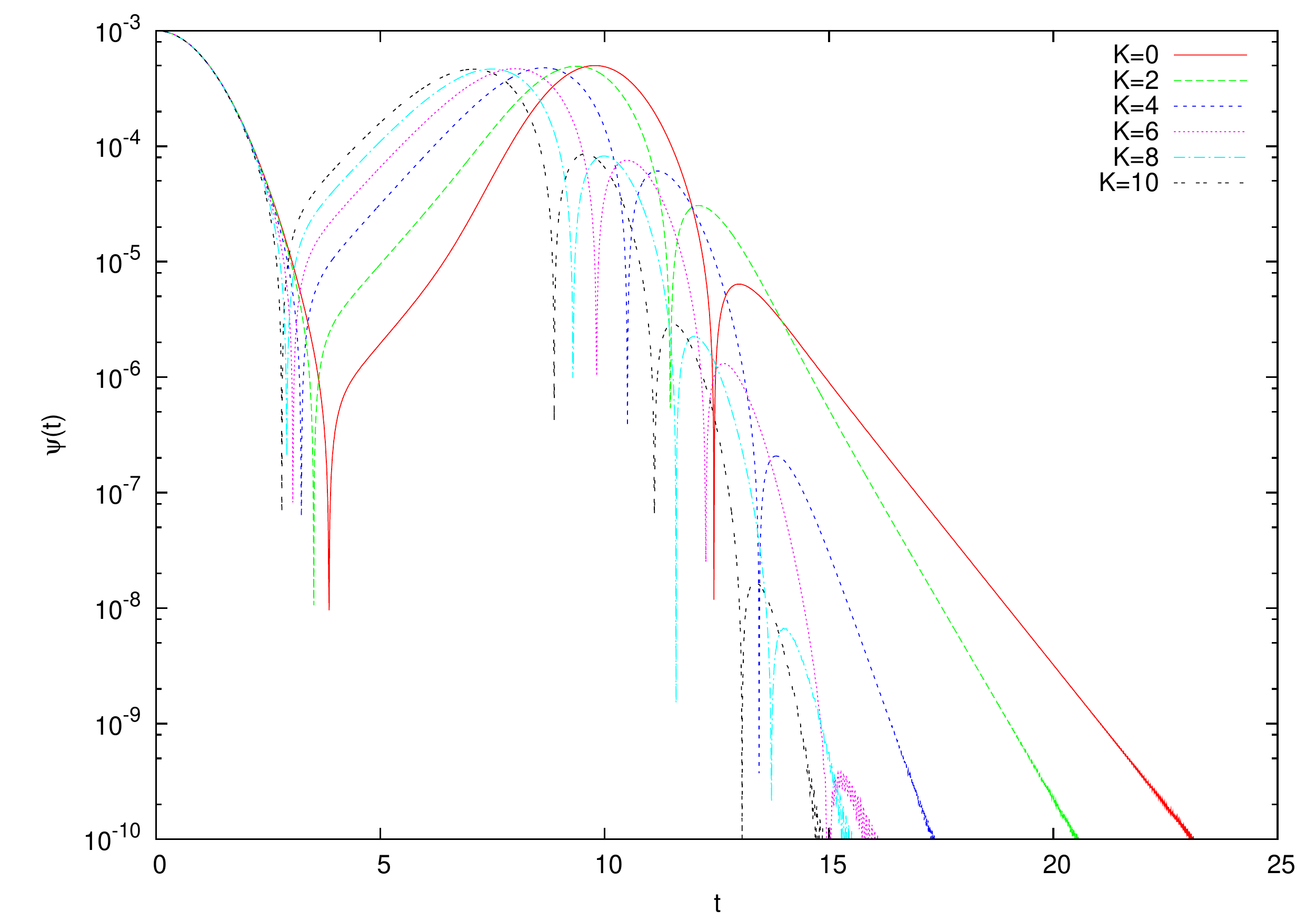}
\caption{Decay of scalar field with mass $m=1$ and $l=1$ varying the
    azimuthal parameter $\kappa$ .}
\label{escalar_varK}
\end{center}
\end{figure}


\begin{figure}[h!]
\begin{center}
\includegraphics[width=7cm, height=9cm,angle=270]{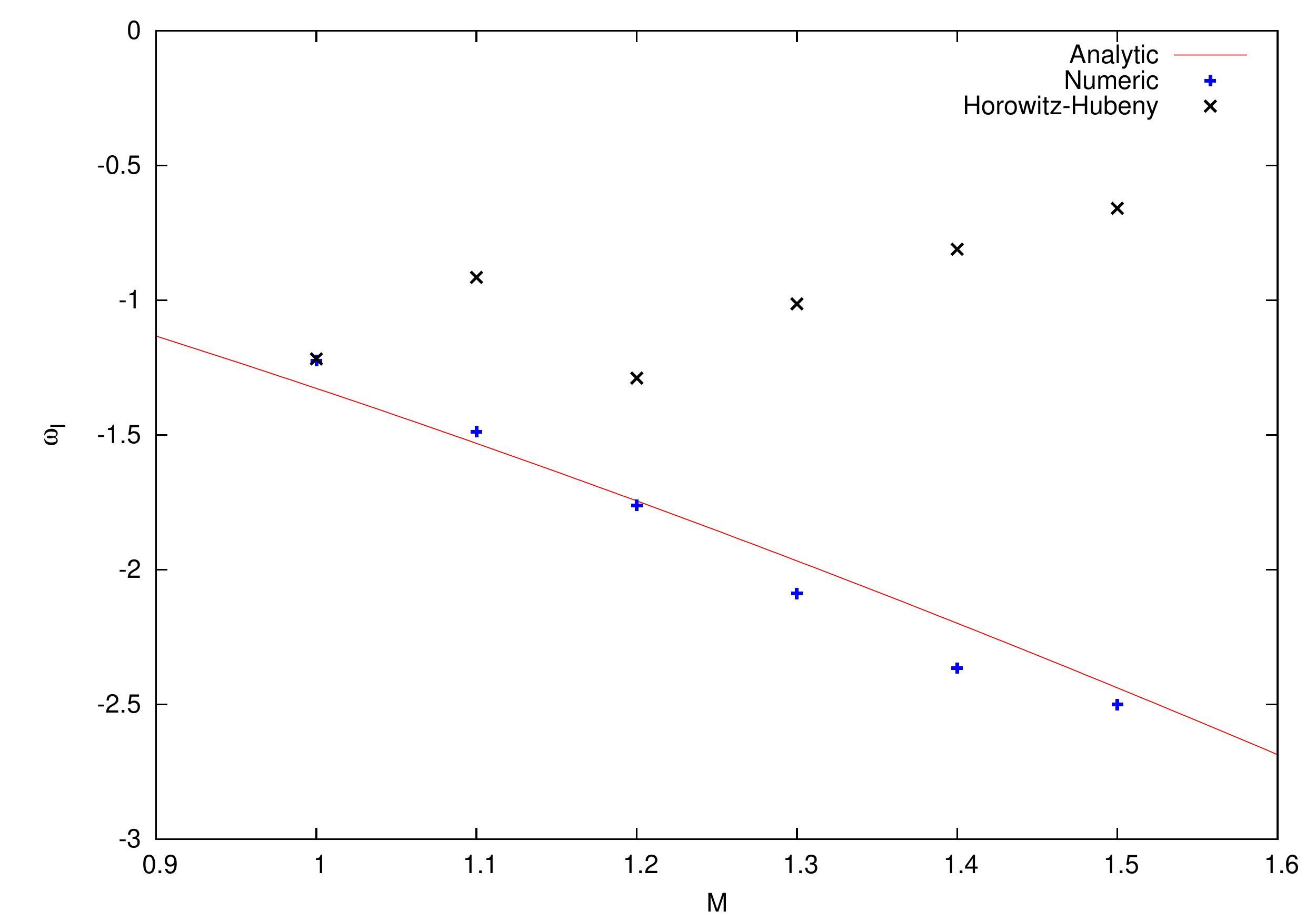}
\caption{Imaginary part of scalar quasinormal frequencies. We
    display the results obtained using different approaches.}
\label{comparacao}
\end{center}
\end{figure}
If $m^2l^2>-1$, we see that $z_-<z_+$. Thus, going back to the
original variable $r$, the roots of the potential are $r=0$ with
double multiplicity, $r=\sqrt{z_-}$ and $r=r_h$ ($r=-\sqrt{z_-}$ and
$r=-r_h$ are excluded as $r>0$). Then, since $r_h$ is the biggest root
and $V(r)$ tends to $\infty$ when $r$ tends to $\infty$, the potential is positive-definite in the region $(r_h,\infty)$. Therefore, the quasinormal modes for $m^2l^2>-1$ are necessarily stable~\cite{hubeny}.

The numerical results regarding the decay of the scalar field are shown in
Figs.\ref{escalar_varM}-\ref{escalar_varK}, and the comparison
between the numerical and analytical results is displayed in
Fig.\ref{comparacao}. Our results reinforce the conclusion already
found analytically; the $z=3$ Lifshitz black hole is stable under
scalar perturbations. Moreover, according to Fig.\ref{comparacao}, the
numerical results have a very good agreement with the analytical
calculation. 

Figure \ref{comparacao} shows that the Horowitz-Hubeny method gives unreliable
results. In~\cite{carlos}, it is argued that the frequencies do not converge as
required by the method, and that may be explained by ill-conditioned
polynomials. However, in this work, the frequencies converge, but they
do not agree with the analytic expression and with the results from
finite difference method. In~\cite{jaqueline}, the authors find cases 
where this method does not work either, and they do so by comparing the 
results with other methods. For instance, they find that the method is 
unreliable for dimensions bigger than 6. Even in the original 
work~\cite{hubeny} the method is unreliable for small black holes, 
and there is no clear explanation for this limitation. In our case the 
asymptotic behavior of the black hole under study might play an important 
role in the convergence of the method. Nevertheless, a general criteria 
for the convergence of the Horowitz-Hubeny method remains an open question.

\bigskip

In the case of the massless spinorial perturbation the superpartner potentials (\ref{potential-weyl}) can be written as

\begin{equation}
V_{\pm}=\frac{1}{l^8}\left[(ml)^2 r^2 (r^2 -r_+ ^2) \pm (ml)r^2 (2r^2-r_+ ^2)\sqrt{r^2-r_+ ^2}\right]\,, 
\end{equation}
and their derivative turns to be
\begin{widetext}
\begin{equation}
V_\pm ' =  \frac{1}{l^8}\left\{ (ml)^2 r (2r^2 -r_+ ^2) \pm (ml)\left[2r(r^2-r_+ ^2)\sqrt{r^2-r_+ ^2} +r^3 \frac{2r^2-r_+ ^2}{\sqrt{r^2-r_+ ^2}}\right]\right\} \,.
\end{equation}
\end{widetext}

\begin{figure}[h!]
\begin{center}
\includegraphics[width=7cm, height=9cm,angle=270]{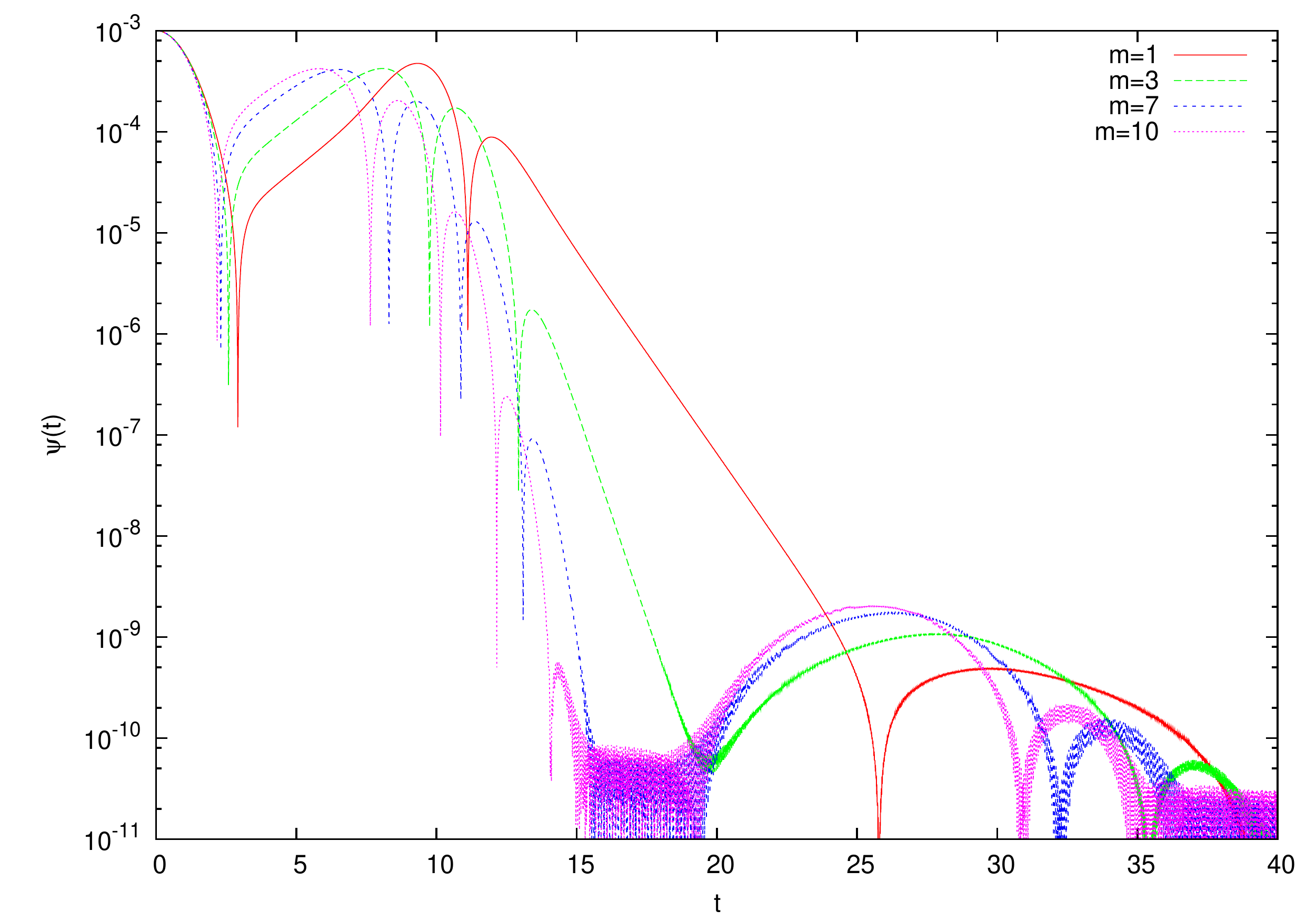}
\caption{Decay of massless spinor with $l=1$ and black hole mass $M=1.0$ for different values of the azimuthal parameter $m$.}
\label{espinorM10}
\end{center}
\end{figure}

\begin{figure}[h!]
\begin{center}
\includegraphics[width=7cm, height=9cm,angle=270]{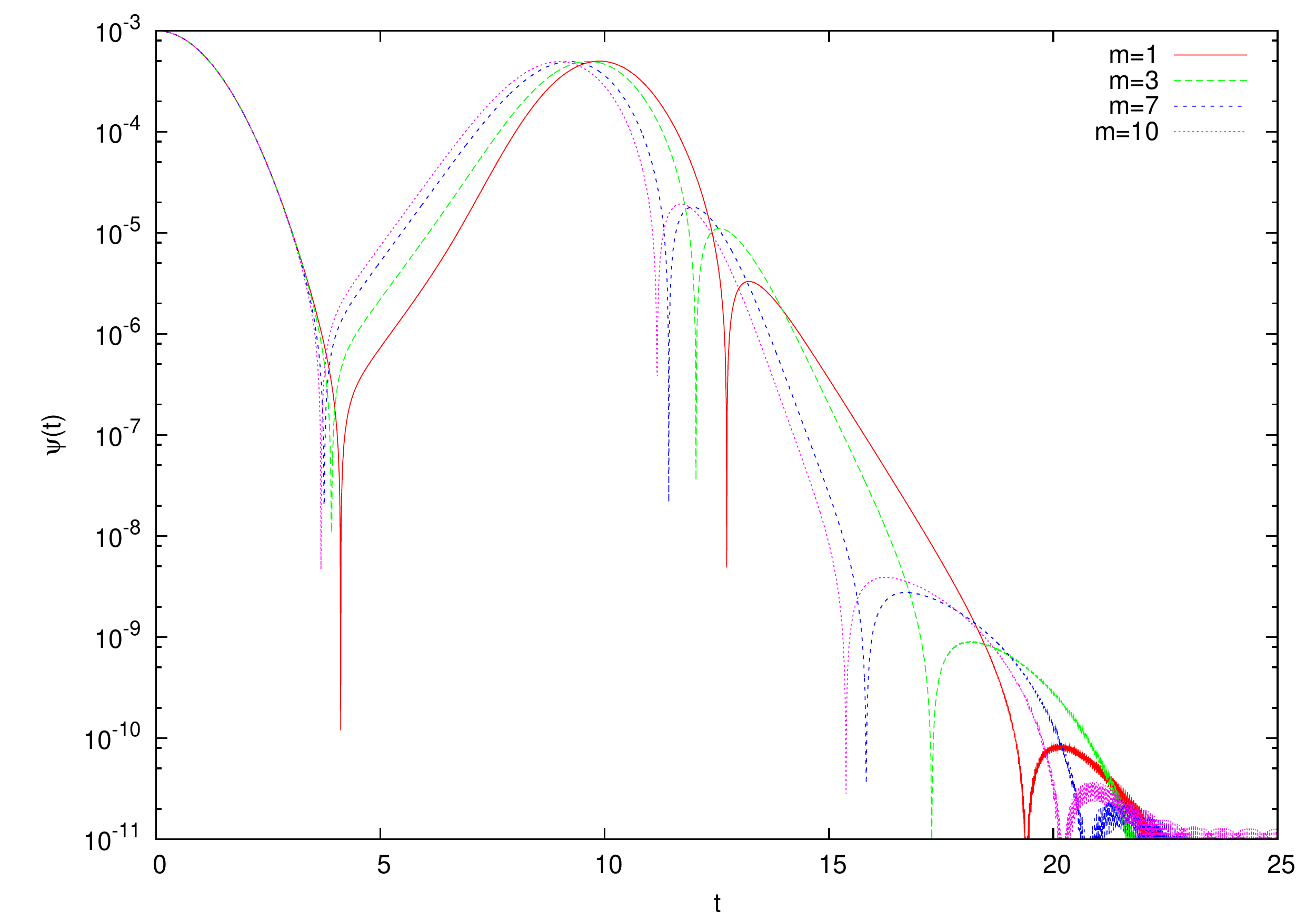}
\caption{Decay of massless spinor with $l=1$ and black hole mass $M=1.5$ for different values of the azimuthal parameter $m$.}
\label{espinorM15}
\end{center}
\end{figure}

We can see that outside the event horizon $V_+$ is positive-definite
if $ml>0$, and $\lim_{r\rightarrow\infty} V_+(r)=-\infty$ if
$ml<0$. Whereas $V_-$ is positive-definite if $ml<0$, and
$\lim_{r\rightarrow\infty} V_-(r)=-\infty$ if $ml>0$. Moreover, we
notice that if $ml=0$, we have a free-particle case. The decaying behavior 
of the massless spinor is given in Figs.\ref{espinorM10}-\ref{espinorM15}. 
Thus, we conclude that the $z=3$ Lifshitz black hole is stable under 
massless spinorial perturbations.


\section{Area Spectrum}

One of the applications of our results for the quasinormal frequencies is the relation they have with the area spectrum of a black hole. According to Maggiore~\cite{maggiore}, the proper physical frequency of the damped harmonic oscillator equivalent to the black hole quasinormal mode is given by
\begin{equation}
\omega_p = \sqrt{\omega_R ^2 + \omega_I ^2}\,,
\end{equation}
where $\omega_R$ and $\omega_I$ stand for the real and imaginary part of the asymptotic QNF, respectively. Thus, using (\ref{asymptotic}) we have
\begin{equation}
\omega_p = 2(\sqrt{6}-2) \frac{M^{3/2}}{l} N\,.
\end{equation}

According to Myung {\it et al.}~\cite{myung}, the Arnowitt-Deser-Misner (ADM) mass of the Lifshitz black hole we are studying is given by
\begin{equation}
{\cal M} = \frac{M^2}{2}\,.
\end{equation}
Applying Maggiore's method, we calculate the adiabatic invariant $I$ as
\begin{equation}
I =\int \frac{d{\cal M}}{\Delta\omega} = \int\frac{M}{\Delta\omega}dM\,,
\end{equation}
where $\Delta\omega$ is the change of proper frequency between two neighboring modes, {\it i.e.}, 
\begin{equation}
\Delta\omega = 2(\sqrt{6}-2) \frac{M^{3/2}}{l} \,.
\end{equation}
Thus,
\begin{equation}\label{invariant}
I = \frac{l M^{1/2}}{(\sqrt{6}-2)}\,,
\end{equation}
which is quantized via Bohr-Sommerfeld quantization in the semiclassical limit. 
Recalling that the horizon area of the black hole is given by $A=2\pi r_+$, with $r_+=l\sqrt{M}$, and using (\ref{invariant}), we arrive at the result,
\begin{equation}\label{areasp}
A = 2\pi (\sqrt{6}-2) n \hbar\,,
\end{equation}
with $n$ an integer number. Therefore, we see that the horizon area for the $z=3$ Lifshitz black hole is quantized and equally spaced.

This result would not be expected for a theory containing higher order curvature corrections since, in general, black hole solutions in such theories do not have a proportional relation between their entropy and area, and consequently, both of them (if any) should not be quantized with an equally spaced spectrum for large quantum numbers~\cite{pap,lanczos,elcio}. However, it was already demonstrated that the $z=3$, three-dimensional Lifshitz black hole has an entropy proportional to its horizon area~\cite{myung,entrop}. Thus, our result (\ref{areasp}) also states that the entropy should be quantized with a spectrum evenly spaced. Nevertheless, we should stress that a generalization of this result for Lifshitz black holes should wait for the calculation of the area spectra of other black holes of such a type. Solely these studies can give a definite answer on this subject.

\section{Concluding Remarks}
We have studied the stability of the three-dimensional Lifshitz black hole
under scalar and spinorial perturbations in the probe limit through the
computation of 
quasinormal modes. In addition, we have found the event horizon area
quantization as an application of the results for quasinormal modes
using Maggiore's prescription. 

Regarding the stability, we have not found unstable quasinormal
modes in the range of parameters that we have considered; all the frequencies 
have a negative imaginary part indicating that
the modes are damped and thus, the perturbations decay, leaving the
system stable against this particular sort of probe fields . 

In the case of a scalar probe field, such results totally agree with the analytical
expressions for the quasinormal frequencies; they show a very large
imaginary part and a very small real part. These modes are almost purely
imaginary.
We have implemented two different numerical methods in order to obtain
the quasinormal frequencies and modes: the finite difference 
and the Horowitz-Hubeny methods. The former allows us to obtain the
temporal behavior of the fields showing all the stages of the
decay, while the latter gives only the frequencies values. As explained in 
section \ref{numerical}, the 
Horowitz-Hubeny method failed in the calculation of the
scalar frequencies as it can be observed in Fig.\ref{comparacao}. 
On the contrary, the finite difference method has a very good 
agreement with the analytical
expression (\ref{freq_analitico}). Apart from the numerical factor,
the asymptotic scalar frequency found in the present work is the same as the one calculated in the
hydrodynamic limit of the scalar perturbations in the context of
gauge-gravity duality\cite{abdallajef}.

Regarding the spinorial perturbation, our numerical results show that 
the probe massless spinor decays and, thus, the $z=3$ Lifshitz black hole is 
stable also under spinorial perturbations.

As a by-product we also obtained the area spectrum of this black hole by 
means of the application of Maggiore's method using 
our results for the scalar asymptotic quasinormal frequencies. Equation (\ref{areasp}) shows that the horizon area is quantized and equally spaced. Furthermore, in light of the conclusions shown in~\cite{myung,entrop}, the corresponding entropy should also have an evenly spaced spectrum. 

Finally, although we have demonstrated the stability of the $z=3$, three-dimensional Lifshitz black hole under scalar and spinor perturbations, we should stress that the definite answer on stability should come from the gravitational perturbations, in particular, from the tensor part of the metric perturbation. It is well known that Einstein gravity in three dimensions has no propagating degrees of freedom, however, the massive versions of the theory, {\it e.g.}, NMG,  allow the propagation of gravitational waves. Albeit this subject deserves further study, the calculation of metric perturbations is a formidable task that is out of the scope of the present paper. The analysis is not dead easy because the perturbation equation is a fourth order differential equation. Thus, some other techniques need to be used together with the traditional QNM analysis~\cite{myung2}. This study will be discussed elsewhere.

\section*{Acknowledgments}

We thank E. Abdalla and A. M. da Silva for enlightening discussions and remarks. 
We also thank E. Papantonopoulos and G. Giribet for reading the manuscript and pointing out useful suggestions, and J. Oliva for interesting comments.
This work was supported by Funda\c c\~ao de Amparo \`a Pesquisa do Estado de S\~ao Paulo {\bf (FAPESP-Brazil)}, State Scholarships Foundation {\bf (IKY-Greece)}, and Consejo Nacional de Investigaciones Cient\'{i}ficas y T\'ecnicas {\bf (CONICET-Argentina)}.


\begin{thebibliography}{99}

\bibitem{rg} 
T.~Regge and J.~A.Wheeler, 
Phys.\ Rev.\ {\bf 108}, 1063-1069 (1957).

\bibitem{zerilli} 
F.~J.~Zerilli,
Phys.\ Rev.\ D {\bf 2}, 2141-2160 (1970).

\bibitem{vivesh} 
C.~V.~Vishveshwara,
Nature {\bf 227}, 936-938 (1970).

\bibitem{press} 
W.~H.~Press,
Astrophys.\ J.\ {\bf 170}, L105-L108 (1971).

\bibitem{chandra} 
S. Chandrasekhar and S. Detweiler, 
Proc.\ R.\ Soc.\ London A {\bf 344}, 441 (1975).

\bibitem{kok} 
K.~D.~Kokkotas, B.~G.~Schmidt, 
Liv.\ Rev.\ Rel.\ {\bf 2}, 2 (1999).

\bibitem{nol} 
H.~P.~Nollert, 
Class.\ Quant.\ Grav.\ {\bf 16}, R159-R216 (1999).

\bibitem{bcs}
E.~Berti, V.~Cardoso, and A.~O.~Starinets, 
Class.\ Quant.\ Grav.\ {\bf 26}, 163001 (2009).


\bibitem{elcio1}
  B.~Wang, C.~-Y.~Lin, E.~Abdalla,
  Phys.\ Lett.\  {\bf B481}, 79-88 (2000).

\bibitem{elcio2}
  B.~Wang, E.~Abdalla, R.~B.~Mann,
  Phys.\ Rev.\  {\bf D65}, 084006 (2002).

\bibitem{cardosoBTZ}
  V.~Cardoso, J.~P.~S.~Lemos,
  Phys.\ Rev.\  {\bf D63}, 124015 (2001).


\bibitem{branas}
  E.~Abdalla, O.~P.~F.~Piedra, J.~de Oliveira,
  Phys.\ Rev.\  {\bf D81}, 064001 (2010).




\bibitem{maldacena}
J.~M.~Maldacena,
Adv.\ Theor.\ Math.\ Phys.\  {\bf 2}, 231 (1998); 
Int.\ J.\ Theor.\ Phys.\  {\bf 38}, 1113 (1999).


\bibitem{birm}
  D.~Birmingham, I.~Sachs, S.~N.~Solodukhin,
  Phys.\ Rev.\ Lett.\  {\bf 88}, 151301 (2002).


\bibitem{hubeny}
G.~T.~Horowitz and V.~E.~Hubeny,
Phys.\ Rev.\ D {\bf 62}, 024027 (2000). 

\bibitem{bek}
J.~D.~Bekenstein, 
Lett.\ Nuovo Cim.\ {\bf 11}, 467 (1974), 
[arXiv:gr-qc/9710076].

\bibitem{hod}
S.~Hod, 
Phys.\ Rev.\ Lett.\ {\bf 81}, 4293 (1998).

\bibitem{kuns}
G.~Kunstatter, 
Phys.\ Rev.\ Lett.\ {\bf 90}, 161301 (2003).

\bibitem{maggiore}
M.~Maggiore,
Phys.\ Rev.\ Lett.\ {\bf 100}, 141301 (2008).


\bibitem{othmet}
A.~J.~M.~Medved, 
Mod.\ Phys.\ Lett.\ A {\bf 24}, 2601 (2009); 
K.~Ropotenko, 
Phys.\ Rev.\ D {\bf 80}, 044022 (2009); 
T.~Padmanabhan and A.~Patel, 
[arXiv:gr-qc/0309053].

\bibitem{pap}
P.~Gonzalez, E.~Papantonopoulos, and J.~Saavedra, 
JHEP {\bf 1008}, 050 (2010).

\bibitem{magg}
S.~W.~Wei, R.~Li, Y.~X.~Liu, and J.~R.~Ren, 
JHEP {\bf 0903}, 076 (2009);
E.~Vagenas, 
JHEP {\bf 0811}, 073 (2008);
A.~J.~M.~Medved, 
Class.\ Quant.\ Grav.\ {\bf 25}, 205014 (2008);
W.~Li, L.~Xu, and J.~Lu, 
Phys.\ Lett.\ B {\bf 676}, 177 (2009);
S.~Fernando, 
Phys.\ Rev.\ D {\bf 79}, 124026 (2009);
S.~W.~Wei and Y.~X.~Liu, 
[arXiv:0906.0908];
D.~Kothawala, T.~Padmanabhan, and S.~Sarkar, 
Phys.\ Rev.\ D {\bf 78}, 104018 (2008);
R.~Banerjee, B.~Majhi, and E.~Vagenas, 
Phys.\ Lett.\ B {\bf 686}, 279-282 (2010);
A.~Lopez-Ortega, 
Phys.\ Lett.\ B {\bf 682}, 85 (2009);
M.~Setare and D.~Momeni, 
Mod.\ Phys.\ Lett.\ A {\bf 26}, 151-159 (2011);
B.~Majhi, 
Phys.\ Lett.\ B {\bf 686}, 49 (2010);
Y.~Kwon and S.~Nam, 
Class.\ Quant.\ Grav.\ {bf 27}, 125007 (2010);
Y.~S.~Myung, 
Phys.\ Lett.\ B {\bf 689},42 (2010);
S.~W.~Wei, Y.~X.~Liu, K.~Yang, and Y.~Zhong, 
Phys.\ Rev.\ D {\bf 81}, 104042 (2010);
M.~R.~Setare, 
Class.\ Quant.\ Grav.\ {\bf 21}, 1453-1458 (2004);
M.~R.~Setare, 
Phys.\ Rev.\ D {\bf 69}, 044016 (2004).


\bibitem{aggh}
E.~Ay\'on-Beato, A.~Garbarz, G.~Giribet, and M.~Hassa\"ine,
Phys.\ Rev.\ D {\bf 80}, 104029 (2009).

\bibitem{unit}
M.~Nakasone and I.~Oda, 
Prog.\ Theor.\ Phys.\ {\bf 121}, 1389 (2009).
S.~Deser, 
Phys.\ Rev.\ Lett.\ {\bf 103}, 101302 (2009). 

\bibitem{ren}
I.~Oda, 
JHEP {\bf 0905}, 064 (2009). 

\bibitem{nmg}
E.~Bergshoeff, O.~Hohm, and P.~Townsend, 
Phys.\ Rev.\ Lett. {\bf 102}, 201301 (2009).

\bibitem{nmg2} 
E.~Bergshoeff, O.~Hohm, and P.~Townsend,
Phys.\ Rev.\ D {\bf 79}, 124042 (2009).

\bibitem{bhnmg}
G.~Clement, 
Class.\ Quant.\ Grav.\ {\bf 26}, 105015 (2009); 
J.~Oliva, D.~Tempo, and R.~Troncoso, 
JHEP {\bf 0907}, 011 (2009);
G.~Giribet, J.~Oliva, D.~Tempo, and R.~Troncoso, 
Phys.\ Rev.\ D {\bf 80}, 124046 (2009); 
A.~Ghodsi, M.~Moghadassi, 
Phys.\ Lett.\ B {\bf 695}, 359 (2011).

\bibitem{kwon}
Y.~Kwon, S.~Nam, J.~D.~Park, S.~H.~Yi, 
Class.\ Quant.\ Grav.\ {\bf 28}, 145006 (2011).


\bibitem{kachru}
S.~Kachru, X.~Liu, and M.~Mulligan, 
Phys.\ Rev.\ D {\bf 78}, 106005 (2008). 

\bibitem{alt}
T.~Azeyanagi, W.~Li, and T.~Takayanagi, 
JHEP {\bf 0906}, 084 (2009).

\bibitem{lifBH1}
R.~Mann, 
JHEP {\bf 06}, 075 (2009). 
 
\bibitem{lifBH2}
G.~Bertoldi, B.~Burrington,and A.~Peet, 
Phys.\ Rev.\ {\bf D80}, 126003 (2009).

\bibitem{lifBH3} 
U.~Danielsson and L.~Thorlacius, 
JHEP {\bf 0903}, 070 (2009).

\bibitem{lifBH4}
K.~Balasubramanian and J.~McGreevy, 
Phys.\ Rev.\ {\bf D80}, 104039 (2009). 

\bibitem{finster1} F., Finster, J. Smoller, S.-T. Yau,
  Phys. Rev. D {\bf{59}}, 104020, (1999).


\bibitem{finster2} F., Finster, N. Kamaran, J. Smoller, S.-T. Yau,
  Adv. Math. Phys. {\bf{7}}, 1, (2003).

\bibitem{cho1}
  H.~T.~Cho, A.~S.~Cornell, J.~Doukas, W.~Naylor,
  Phys.\ Rev.\  {\bf D75}, 104005 (2007).


\bibitem{cho2}
  H.~T. ~Cho, 
  Phys.\ Rev.\  {\bf D68}, 024003 (2003);
  J.~Jing, 
  Phys.\ Rev.\ {\bf D71}, 124006 (2005). 

\bibitem{owenjef}
  O.~P.~F.~Piedra, J.~de Oliveira,
  Class.\ Quant.\ Grav.\  {\bf 28}, 085023 (2011).

\bibitem{kazakov}
A.~Ya.~Kazakov, 
J.\ Phys.\ A: Math.\ Gen.\ {\bf 39}, 2339 (2006).


\bibitem{cardoso}
V.~Cardoso and J.~P.~S.~Lemos,
Phys.\ Rev.\  D {\bf 63}, 124015 (2001). 

\bibitem{myung}
Y.~S.~Myung, Y.~W.~Kim, and Y.~J.~Park, 
Eur.\ Phys.\ J.\ C{\bf 70}, 335-340 (2010).

\bibitem{lanczos}
D.~Kothawala, T.~Padmanabhan, S.~Sarkar, 
Phys.\ Rev.\ D {\bf 78}, 104018 (2008).

\bibitem{elcio}
E.~Abdalla, L.~A.~Correa-Borbonet, 
Phys.\ Rev.\ D {\bf 65}, 124011 (2002); 
E.~Abdalla, L.~A.~Correa-Borbonet, 
Mod.\ Phys.\ Lett.\ A {\bf 16}, 2495-2504 (2001). 

\bibitem{entrop}
R.-G.~Cai, Y.~Liu, Y.-W.~Sun, 
JHEP {\bf 0910}, 080 (2009); 
O.~Hohm, E.~Tonni, 
JHEP {\bf 1004}, 093 (2010); 
H.~Gonzalez, D.~Tempo, R.~Troncoso, 
JHEP {\bf 11}, 066 (2011). 

\bibitem{abdallajef}
E.~Abdalla, J.~de Oliveira, A.~Lima-Santos and A.~B.~Pavan,
[arXiv:1108.6283].

\bibitem{carlos}
  E.~Abdalla, C.~E.~Pellicer, J.~de Oliveira, A.~B.~Pavan,
  Phys.\ Rev.\  {\bf D82}, 124033 (2010).

\bibitem{jaqueline}
  J.~Morgan, V.~Cardoso, A.~S.~Miranda, C.~Molina, V.~T.~Zanchin,
  JHEP {\bf 0909}, 117 (2009).

\bibitem{myung2}
  Y.~S.~Myung, Y.-W.~Kim, T.~Moon, Y.-J.~Park, 
  Phys.\ Rev.\ D {\bf 84}, 024044 (2011).


\end{thebibliography}
\end{document}